\documentstyle[12pt]{article}

\def\d{\partial}

\def\a{\alpha}
\def\th{\theta}
\def\g{\gamma}
\def\G{\Gamma}

\def\N{\nabla}

\def\de{\delta}
\def\P{\Phi}
\def\f{\varphi}

\def\r{\rho}

\def\L{\Lambda}

\def\s{\sigma}

\def\F{\Psi}
\def\e{\epsilon}

\def\t{\tilde}
\def\implies{\Rightarrow}
\def\is{\equiv}

\def\frak#1#2{ \textstyle{#1\over #2}}

\def\half{\frac{1}{2}}

\newcommand{\nc}{\newcommand}
\nc{\beq}{\begin{equation}}
\nc{\eeq}[1]{\label{#1}\end{equation}}
\nc{\ber}{\begin{eqnarray}}
\nc{\eer}[1]{\label{#1}\end{eqnarray}}
\nc{\pek}[1]{\cite{#1}}
\nc{\enr}[1]{(\ref{#1})}
\nc{\kal}[1]{{\cal{#1}}}
\nc{\dott}{\!\cdot\!}
\newwrite\ffile\global\newcount\figno \global\figno=1

\def\writedef#1{}

\def\figin{\epsfcheck\figin}\def\figins{\epsfcheck\figins}
\def\epsfcheck{\ifx\epsfbox\UnDeFiNeD
\message{(NO epsf.tex, FIGURES WILL BE IGNORED)}
\gdef\figin##1{\vskip2in}\gdef\figins##1{\hskip.5in}
instead
\else\message{(FIGURES WILL BE INCLUDED)}%
\gdef\figin##1{##1}\gdef\figins##1{##1}\fi}

\def\figinsert{}
\def\ifig#1#2#3{\xdef#1{fig.~\the\figno}
\writedef{#1\leftbracket fig.\noexpand~\the\figno}%
\figinsert\figin{\centerline{#3}}\medskip\centerline{\vbox{\baselineskip12pt
\advance\hsize by -1truein\center\footnotesize{  Fig.~\the\figno.} #2}}
\bigskip\endinsert\global\advance\figno by1}
\def\endinsert{}

\begin{document}

\title{\large{\bf World-Volume Locally Supersymmetric Born-Infeld Actions}}

\author{ Bj\"orn Brinne\thanks{brinne@physto.se}\,\,$^a$,\,\,\,
 Svend E. Hjelmeland\thanks{s.e.hjelmeland@fys.uio.no}\,\,$^b$ and
 Ulf Lindstr\"om\thanks{ul@physto.se and ulfl@boson.uio.no}\,\,$^{a,b}$ \\
\, \\
{\small{$^{a}$ITP, University of
Stockholm,}}\\
{\small{ Box 6730, S-11385 Stockholm, Sweden}}\\
\, \\
{\small{$^b$Department of Physics, University of Oslo}}\\
{\small{P.O.Box 1048 Blindern, N-0316 Oslo, Norway}}\\}

\date{}

\maketitle

\begin{picture}(0,0)(0,0)
\put(390,310){USITP-99-02}
\put(390,295){OSLO-TP 3-99}
\put(390,280){hep-th/9904175}
\end{picture}
\vspace{-24pt}

\begin{abstract}
We derive manifestly locally supersymmetric extensions of
the Born-Infeld action with $p=2$. The construction is based on
a first order bosonic action for $Dp$-branes with a generalized Weyl
invariance.

\end{abstract}

\newpage

\section{Introduction}

In string theory, the Born-Infeld action \pek{boin} arises as (part of) the
low energy effective action of open superstring theory \pek{frtse, tsey,
acny, clny}, and as (the bosonic part of) the effective actions for
$Dp$-branes in type $IIA$ and $IIB$ theories \pek{leig}. Various 
supersymmetrizations of the BI action have been discussed.  Global 
supersymmetrization of the ambient spacetime variables is discussed 
in \pek{apsc, cgnw, cgnsw, bsto, beto}. This gives the analogue of 
the Green-Schwarz string action and thus also includes local 
$\kappa$-symmetry. Global world volume supersymmetrization is 
discussed in component form for $p=3$ in \pek{depu, cefe, Karlhede:1987bg}. For 
certain $p$'s global superspace formulation of world volume 
supersymmetry also exist \pek{baga, rotse, ket1, ket2, Ivanov:1999fw}. 

In this paper we try to combine the advantages of having a
geometrical first order formulation involving an auxiliary metric with the
advantage of having supersymmetry manifest. This requires a
world-volume superspace supergravity extension of the Born-Infeld action.
As the form of superspace supergravity is highly dimension dependent we do
not expect to find a general prescription. Below we restrict ourselves to
$p=2$. This case is also of interest since $D=3$ supergravity is
non-dynamical and one might hope to retain some simplicity 
even in the presence of the supergravity action.

We base our construction on a bosonic first order action for arbitrary
$Dp$-branes. This action has a generalized type of Weyl invariance which
reduces the number of degrees of freedom in the auxiliary second rank
tensor field. It would have been nice to be able to extend this to a
super-Weyl invariance for the locally supersymmetric theory, and indeed we
find a super-Weyl invariant candidate action. Unfortunately it leads to
unwanted bosonic terms, however.

The plan of the paper is as follows: In Sec.$2$ we review some background
mainly concerning the spinning membrane. Sec.$3$ introduces the new
bosonic Born-Infeld action and Sec.$4$ presents $D=3$ superspace along
with our superspace actions. In Sec.$5$ we give the reduction of the
superfields to components preparing for Sec.$6$ where the component form
of the actions are given. Sec.$7$, finally, contains our conclusions.

\section{Background}

In \pek{ul} the following general Weyl invariant action for $p$-branes
was presented:
\beq
I_W=\int d^{p+1}\xi\sqrt{-g}(g^{mn}\g_{mn})^{{p+1}\over 2},
\eeq{iw}
where $X^\Lambda,\quad \Lambda=0....,D-1$ are coordinates
in the $D$-dimensional (target) space-time,
$\xi^m,\quad m=0,...,p$ coordinatize the $p$-brane world volume,
$\g_{mn}=\d_mX^\Lambda\d_nX^\Omega G_{\Lambda\Omega}$ is the world-volume metric induced
from the space-time metric $ G_{\Lambda\Omega}$ and $g$ is the determinant of the
auxiliary metric $g_{mn}$.

For $p=1$ the action \enr{iw} agrees
with the usual string action with an auxiliary metric $g_{mn}$ \pek{bdh,dz}.
Coupling it to $2D$ supergravity in superspace leads to the spinning string:
\beq
I=\int d^2\xi d^2\th E^{-1}\N^\a X\N_\a X,
\eeq{ss}
where all fields are superfields, $E$ is the (super-)determinant of the
supervielbein, $\N_\a$ are the spinorial covariant derivatives and we have
suppressed the ambient space-time indices on X.

For $p=2$ the action \enr{iw} agrees with the standard cosmological term action
\pek{ht}
\beq
I=\int d^3\xi\sqrt{-g}(g^{mn}\g_{mn}-1).
\eeq{ca}
After eliminating $g_{mn}$ they both give the Nambu-Goto type action
representing the volume of the world-volume. However, whereas \enr{ca} is
impossible to couple to $3D$-supergravity \pek{bst}, this is not so for the
$p=2$
version of the action \enr{iw}. In \pek{ulmr} it was shown that there exist (at
least) two different supersymmetrizations of \enr{iw} when $p=2$. They read
\beq
I_1=\int d^3\xi d^2\th E^{-1}(\N^\a X\N_\a X)(\N^{\beta\g} X\N_{\beta\g}
X)^{1/2},
\eeq{sm1}
and
\beq
I_2=\int d^3\xi d^2\th
 E^{-1}(i\N^\a X\N_\a^\g X)(i\N^\beta X\N_{\beta\g} X)
(\N^{\s\r}X\N_{\s\r}X)^{-1/2},
\eeq{sm2}
where again all fields are superfields and we have introduced $3D$-superspace
supergravity, to be described in more detail below. In none of the actions
\enr{sm1} and \enr{sm2} does the bosonic Weyl invariance extend to super-Weyl
invariance. However, in \pek{ulmr} it was found that the combination
$I_1-{2\over
3}I_2$ is in fact super-Weyl invariant.

\section{Born-Infeld actions}

The Born-Infeld action
\beq
I^{BI}_{1}=T_p\int d^{p+1}\xi \sqrt{-\det(\g_{mn}+\kal{F}_{mn})}
\eeq{bi}
is a direct generalization of the Nambu-Goto type action for $p$-branes by
inclusion of the two-form field $\kal{F}_{mn}$. In the context of $D$-branes
the brane tension $T_p$ is related to the fundamental string tension $(\a
')^{-1}$ and the string coupling constant $g_s$, and
\beq
\kal{F}_{mn}\is B_{mn}+2\pi \a 'F_{mn},
\eeq{f}
with $F=dA$ the $U(1)$ field strength for the world-volume field $A_m$ and
$B_{mn}$ the pull-back to the world-volume of the Kalb-Ramond antisymmetric
tensor field. (There is also a multiplicative exponent of a dilaton field
in the action which will play no role in our considerations. We set it to
$1$ w.l.o.g.). In analogy to the
$p$-brane case there is a first order action of the cosmological term type for
the action
\enr{bi},
\pek{ul}. It is given by (c.f. \enr{ca}),
\beq
I^{BI}_{2}=T_p\int
d^{p+1}\xi\sqrt{-s}\left[ s^{mn}(\g_{mn}+\kal{F}_{mn})+(p-1)\right] ,
\eeq{sa1}
where $s^{mn}(\xi)$ is a general tensor (it has no symmetry) and $s\is
\det{s_{mn}}$. This form of the action has been used in \pek{ulrvu} as the
starting point for a discussion of the strong coupling limit of $D$-branes. The
discussion was extended to include spin in that limit in \pek{hgul}. However,
the same objections to a direct local
supersymmetrization that were raised in \pek{bst} apply to the action \enr{sa1}
for
$p=2$, since then
\enr{sa1} becomes \enr{ca} for $\kal{F}=0=s^{[mn]}$. We thus have to look
for an
action analogous to \enr{iw} to use as a starting point. For $p=3$ this was 
given in \pek{ul} and for general $p$ it is
\beq
I^{BI}_{3}=T_p\int d^{p+1}\xi \sqrt{-s}\left[
s^{mn}(\g_{mn}+\kal{F}_{mn})\right]^{{p+1}\over 2}.
\eeq{sa2}
It is easy to see that eliminating $s_{mn}$ the action \enr{bi} is
recovered\footnote{We are being careless with normalisation. For exact
equivalence we should specify the $p$-dependent numerical factors in front of
the actions.}. Furthermore we note that the action
\enr{sa2} has a (generalized) Weyl invariance and that it reduces to the
action
\enr{iw} in the limit
$\kal{F}=0=s^{[mn]}$.

We now specialize to $p=2$. Since our goal is a locally supersymmetric
superspace action of the \enr{sm1} or \enr{sm2} type, we must reformulate
\enr{sa2} in terms of vielbeins. As a first step in that direction we separate
the symmetric and antisymmetric part in $s^{mn}$ according to
\beq
s^{mn}\is g^{mn}+\e^{mnp}h_p/\sqrt{-g}.
\eeq{gdef}
It follows that $\sqrt{-s}=\sqrt{-g}/\sqrt{1-h^2}$, and the action
\enr{sa2} becomes, (properly normalized for $p=2$ and with $T_p$ set to
one),
\beq
{1\over{3\sqrt{3}}}\int d^3\xi \sqrt{{{-g}\over{1-h^2}}}\left[
g^{mn}\g_{mn}+\e^{mnp}h_p\kal{F}_{mn}/\sqrt{-g}\right]^{3/2},
\eeq{gha}
where $h^2\is g^{mn}h_mh_n$. Introducing vielbeins $e_a\is e_a^m\d_m$ and their
inverses $e^a\is e^a_md\xi^m$ according to
\beq
g_{mn}=e^a_me^b_n\eta_{ab}, \quad e_m^ae_a^n= \de^n_m,\quad e^{-1}\is
\det(e^a_m),
\eeq{vb}
where $a=0,1,2$ are tangent space indices and $\eta_{ab}$ is the Minkowski
metric, we may rewrite the action \enr{gha} as follows:
\beq
{1\over{3\sqrt{3}}}\int d^3\xi {e^{-1}\over\sqrt{1-h^2}}\left[
e_aXe^aX+e^m_ae^n_b\e^{abc}h_c\kal{F}_{mn}\right]^{3/2}.
\eeq{eha}
The independent fields are now $X^\Lambda, A_m, e_a^m$ and $h_a$. As a check,
we have verified that
\enr{eha} is indeed equivalent to the $3D$ Born-Infeld action \enr{bi}. The
Weyl invariance is now of the usual type, i.e., a rescaling of $e^a_m$ with
$h_a$ inert. This form is suitable for supersymmetrization.

\section{Superspace}

We will use the $3D$ superspace conventions of \pek{book}. The supergravity
algebra is given by
\ber
&&\{\N_\a,\N_\beta\}=2i\N_{\a\beta},\quad
[\N_\a,\N_{\beta\g}]=C_{\a(\beta}U_{\g)},\cr
&&U_\a=-iR\N_\a+i{2\over 3}(\N_\beta
R)M_\a^{\ \beta} +iG_{\a\beta}^{\ \ \g}M_\g^{\ \beta},
\eer{alg}
where $M$ are Lorentz generators and $R$ and $G_{\a\beta\g}$ are the basic
superfields that solve the Bianchi identities ($G_{\a\beta\g}$ is completely
symmetric in all three spinor indices.). In this notation a vector
index is represented by a symmetrized pair of spinor indices. The
covariant derivatives have the usual structure $\N_A=E_A^MD_M+\P_A\cdot M$
with $E_A^M$ being the supervielbein, $D_M$ the flat superspace covariant
derivatives and $\P_A$ the connection. The matter
superfields that we will consider are $X^\Lambda(\xi ,\th)$ whose $\th$-independent
part is the space-time coordinate and  $H_\a$ which is a spinor superfield
whose
covariant derivative $H_{\a\beta}=\frac{1}{2}\N_{(\a}H_{\beta)}$ has 
the bosonic vector
$h_a$ as lowest component. Finally, the dual of the Maxwell
field-strength\footnote{From now on we put the background $B$-field to
zero.} is
the lowest component of
$F_{\a\beta}=\frac{1}{2}\N_{(\a}W_{\beta)}$, with
$W_\a$ the electromagnetic spinor potential.

With the above ingredients we immediately write down two possible
generalizations of \enr{sm1} and \enr{sm2}:
\beq
I_1=\int d^3\xi d^2\th E^{-1}(1-H^2)^{-1/2}(\N^\a X\N_\a X
-H^\a W_\a){\cal S}^{1/2},
\eeq{sbi1}
and
\beq
I_2=\int d^3\xi d^2\th
 E^{-1}(1-H^2)^{-1/2}\f^\a\f_\a {\cal S}^{-1/2},
\eeq{sbi2}
where
\ber
&&{\cal S}\is (\N^{\beta\g} X\N_{\beta\g}X+H^{\a\beta}F_{\a\beta}), \cr
&&\f_\g\is
(i\N^\a X\N_{\a\g} X+H_\g^{\ \a}W_\a), \quad H^2\is H^{\a\beta}H_{\a\beta}.
\eer{fdef}
These actions reduce to \enr{sm1} and \enr{sm2} when $H_\a=0$, and
in the next section we show that the bosonic parts of both \enr{sbi1} and
\enr{sbi2} are equivalent to the action \enr{eha}. Before going to components
let us look at (possible) super-Weyl invariance \pek{ht2}. Under an
infinitesimal super Weyl transformation with parameter $\L$ the superfields
transform as follows:
\ber
\de E_\a&=&\L E_\a, \quad \de E_{\a\beta}=2\L E_{\a\beta}
-i(E_{(\a}\L ) E_{\beta)}
\implies \de E^{-1}=-4\L E^{-1},\cr
\de H_\a&=&-\L H_\a, \quad \de H_{\a\beta}=0,\cr
\de W_\a&=&3\L W_\a, \quad \de F_{\a\beta}=4\L F_{\a\beta}+2(E_{(\a}\L )
W_{\beta)},
\eer{swt}
where we have used that \pek{book}
\beq
W_\a\is\N^\beta\N_\a\G_\beta+2R\G_\a,
\eeq{wde}
and that
\ber
\de \G_\a = \L\G_\a, \quad \de R= 2\L R -2\N^2\L, \quad \de \P_{\a\beta\g}=
-(E_{(\g}\L ) C_{\beta)\a}+\L\P_{\a\beta\g}.
\eer{swt2}
The transformation of $H_\a$ was determined from the requirement that $\de
h_a=0$, in agreement with the discussion in Sec.3.

The difficulty in finding an invariant action is entirely due to the
inhomogeneous parts of the transformations in \enr{swt}. We find the following
super-Weyl invariant combination of the action in \enr{sbi2} and a 
modification of the action in \enr{sbi1}:
\beq
I_W=\t I_1-\frak{2}{3}I_2,
\eeq{swa}
where $\t I_1$ indicates that we have made the replacement $(\N^\a X\N_\a X-H^\a
W_\a)\to (\N^\a X\N_\a X)$ in the Lagrangian. As will be seen in Sec.6, the
action $I_W$ in \enr{swa} has a bosonic part that differs from  the action in
\enr{eha}, so it is not a supersymmetrization of that. It does however reduce to  
the super-Weyl invariant combination of
\enr{sm1} and \enr{sm2} when $H_\a=0$.

\section{Components}

There is a systematic procedure for deriving the components, the component 
action and the component local supersymmetry transformations of 
a theory in superspace. It is described in \pek{book} for 
$D=4$ supergravity actions. 
For the case at hand we use the definitions and 
results in \pek{ulmr} to which we add those pertaining to the
$H_\a$ and $W_\a$ fields. 
We define
\ber
X^\Lambda|\equiv A^\Lambda,
\ \ \ \nabla_{\alpha}X^\Lambda|\equiv\chi_{\alpha}^\Lambda,
\ \ \ \nabla^{2} X^\Lambda|\equiv\frac{1}{2}\nabla^{\alpha}
\nabla_{\alpha}X^\Lambda|\equiv{\cal T}^\Lambda,
\nonumber \\
\nabla_{\alpha}|\equiv\partial_{\alpha},
\ \ \ \nabla_{\alpha\beta}|\equiv{\cal D}_{\alpha\beta}+iSM_{\alpha\beta}
+\Psi_{\alpha\beta}^{\ \ \ \gamma}\nabla_{\gamma}|, \nonumber \\
R|\equiv S,
\eer{xdef}
where $|$ denotes ``the $\theta$ independent part of''.
Then the superspace-component relations involving the matter superfield
become (suppressing the space-time indices)
\begin{eqnarray}
\nabla_{\alpha\beta}X|&=&{\cal D}_{\alpha\beta}A
+\Psi_{\alpha\beta}^{\ \ \ \gamma}\chi_{\gamma}
\equiv\hat{\nabla}_{\alpha\beta}A, 
\nonumber \\
\nabla_{\alpha}\nabla_{\beta}X|&=&i\hat{\nabla}_{\alpha\beta}A
-C_{\alpha\beta}{\cal T},
\nonumber \\
\nabla_{\alpha}\nabla_{\beta\gamma}X|&=&{\cal D}_{\beta\gamma}\chi_{\alpha}+
\Psi_{\beta\gamma}^{\ \ \ \delta}(i\hat{\nabla}_{\delta\alpha}A
-C_{\delta\alpha}{\cal T})
-\frac{1}{2}iC_{\alpha(\beta}\chi_{\gamma)}S \nonumber \\
&\equiv&\hat{\nabla}_{\beta\gamma}\chi_{\alpha}
-\frac{1}{2}iC_{\alpha(\beta}\chi_{\gamma)}S, \nonumber \\ 
\nonumber \\
\nabla^{2}\nabla_{\alpha}X|&=&i\hat{\nabla}_{\beta\alpha}\chi^{\beta}
-\frac{1}{2}S\chi_{\alpha},
\nonumber \\
\nabla_{\alpha}\nabla^{2}X|&=&-i\hat{\nabla}_{\beta\alpha}\chi^{\beta}
-\frac{1}{2}S\chi_{\alpha},
\nonumber \\
\nabla^{2}\nabla_{\alpha\beta}X|&=&{\cal D}_{\alpha\beta}{\cal T}
-\Psi_{\alpha\beta}^{\ \ \ \delta}
(i\hat{\nabla}_{\delta\sigma}\chi^{\sigma}+\frac{1}{2}S\chi_{\delta})
+i(\frac{2}{3}\chi_{(\alpha}\eta_{\beta)}
-{\cal G}_{\alpha\beta}^{\ \ \ \sigma}\chi_{\sigma}) \nonumber \\
&\equiv&\hat{\nabla}_{\alpha\beta}{\cal T}
+i(\frac{2}{3}\chi_{(\alpha}\eta_{\beta)}
-{\cal G}_{\alpha\beta}^{\ \ \ \sigma}\chi_{\sigma})-\half S\Psi_{\alpha\beta}{}^\delta\chi_\delta,
\end{eqnarray}
where
\begin{eqnarray}
{\cal G}_{\alpha\beta\gamma}\equiv G_{\alpha\beta\gamma}|&=&
\frac{1}{6}({\cal D}_{\delta(\alpha}\Psi^{\delta}_{\ \beta\gamma)}
-t_{\delta(\alpha\ \beta}^{\ \ \ \delta\ \ \epsilon\lambda}
\Psi_{|\epsilon\lambda|\gamma)}+\frac{1}{2}iS\Psi_{(\alpha\beta\gamma)}),
\nonumber \\
\eta_{\alpha}\equiv\nabla_{\alpha}R|&=&
-\frac{1}{2}({\cal D}_{\delta(\alpha}\Psi_{\gamma)}^{\ \ \delta\gamma}
-t_{\delta(\alpha\ \gamma)}^{\ \ \ \delta\ \ \epsilon\lambda}
\Psi_{\epsilon\lambda}^{\ \ \gamma})+iS\Psi_{\alpha\beta}^{\ \ \ \beta},
\nonumber \\
t_{\alpha\beta\gamma\delta}^{\ \ \ \ \ \epsilon\lambda}&=&
-\frac{1}{2}i(C_{\alpha\gamma}\Psi_{\sigma(\beta}^{\ \ \ \ \epsilon}
\Psi^{\sigma\ \ \lambda}_{\ \delta)}
+C_{\beta\delta}\Psi_{\sigma(\alpha}^{\ \ \ \ \epsilon}
\Psi_{\gamma)}^{\ \ \sigma\lambda}).
\end{eqnarray}
Our vielbein components are defined by
$\nabla_\alpha |$ and $\nabla_{\alpha\beta} |$ in \enr{xdef} to be\footnote{As seen from
the component of the spinorial derivative, we use a Wess-Zumino gauge.}
\begin{equation}
E_\alpha{}^\mu | = \delta_\alpha{}^\mu,\quad
E_\alpha{}^{\mu\nu} | = 0,\quad
E_{\alpha\beta}{}^\mu | = \Psi_{\alpha\beta}{}^\mu,\quad
E_{\alpha\beta}{}^{\mu\nu}| =
  e_{\alpha\beta}{}^{\mu\nu} \equiv e_{\alpha\beta}{}^m
\end{equation}
The components of the prepotential are
\begin{eqnarray}
W_{\alpha}|&\is&\lambda_{\alpha}, \nonumber \\
F_{\alpha\beta}|
&=&\frac{1}{2}\nabla_{(\alpha}W_{\beta)}|=f_{\alpha\beta}, 
\nonumber \\
\nabla_{\alpha}F_{\beta\gamma}|&=&\sigma_{\alpha\beta\gamma}
+SC_{\alpha(\beta}\lambda_{\gamma)}, \nonumber \\
\sigma_{\alpha\beta\gamma}
&\equiv&i\left[{\cal D}_{\alpha(\beta}\lambda_{\gamma)}
-{\cal D}_{\beta\gamma}\lambda_{\alpha}
-f_{\delta\alpha}\Psi_{\beta\gamma}^{\ \ \ \delta}
+f_{\delta(\beta}\Psi_{\gamma)\alpha}^{\ \ \ \delta}\right],
\nonumber \\
\nabla^{2}F_{\beta\gamma}|&\equiv&\sigma_{\beta\gamma}
-\frac{i}{2}S\Psi^\rho{}_{(\beta\gamma)}\lambda_{\rho}
-\frac{i}{2}S\Psi^{\rho}_{\ (\beta|\rho|}\lambda_{\gamma)},
 \nonumber \\
\sigma_{\beta\gamma}&\equiv&i\left[-\frac{i}{3}\eta_{(\beta}\lambda_{\gamma)}
-2{\cal G}_{\beta\gamma}^{\ \ \ \rho}\lambda_{\rho}
+\frac{1}{2}{\cal D}_{\rho(\beta}f_{\gamma)}^{\ \ \rho}
+\frac{1}{2}\Psi_{\rho(\beta}^{\ \ \ \lambda}
\sigma_{|\lambda|\gamma)}^{\ \ \ \ \rho}\right].
\end{eqnarray}
We define the components of the spinor superfield as follows
\begin{eqnarray}
H_\alpha | &\is& h_\alpha, \ \ \ \ \nonumber \\
\nabla_{\alpha}H_{\beta}|&\is&(H_{\alpha\beta}+C_{\alpha\beta}B)|
 \is h_{\alpha\beta}+C_{\alpha\beta}b, \nonumber \\
\nabla^2 H_\alpha | &\is& \mu_\alpha + \half S h_\alpha, \nonumber \\
\nabla_\alpha H_{\beta\gamma} | &\equiv& \phi_{\alpha\beta\gamma} = 
\frac{i}{2}{\cal D}_{\alpha(\beta}h_{\gamma)} - \half C_{\alpha(\beta}{\mu}_{\gamma)} +
\frac{i}{2}\Psi_{\alpha(\beta}{}^\rho h_{\gamma)\rho} + \frac{i}{2}
\Psi_{\alpha(\beta\gamma)}b , \nonumber \\
\nabla^2 H_{\alpha\beta} | &\equiv& \phi_{\alpha\beta} =
\frac{i}{2}{\cal D}_{\gamma(\alpha}h^\gamma{}_{\beta)}-i{\cal D}_{\alpha\beta}b
+\frac{1}{3}\eta_{(\alpha}h_{\beta)} - 2{\cal G}_{\alpha\beta\gamma}h^\gamma + 
\frac{i}{2} \Psi_{\gamma(\alpha}{}^\rho \phi_{|\rho|}{}^\gamma{}_{\beta)} + \nonumber \\
&&+ \half \Psi_{\alpha\beta}{}^\rho {\cal D}_{\rho\sigma}h^\sigma+
\half\Psi_{\alpha\beta}{}^\rho \Psi_{\rho\gamma}{}^\delta h_\delta{}^\gamma +
\frac{i}{2}\Psi_{\alpha\beta}{}^\rho\Psi_{\gamma\rho}{}^\gamma b
-iS\Psi_{\alpha\beta}{}^\rho h_\rho, \nonumber \\ 
\nabla_\gamma B | &\is& b_\g= \half \mu_\gamma+\frac{i}{2}{\cal
D}_{\gamma\alpha}h^\alpha + Sh_\gamma  +
\frac{i}{2}\Psi_{\gamma\alpha}{}^\rho h_\rho{}^\alpha
+\frac{i}{2}\Psi_{\alpha\gamma}{}^\alpha b, \nonumber \\
\nabla^2 B | &=& \frac{i}{2}{\cal D}_{\alpha\beta}h^{\alpha\beta}
+\frac{i}{2}\Psi_{\alpha\beta}{}^\gamma\phi_\gamma{}^{\alpha\beta}-2Sb-\eta_\beta
h^\beta. 
\eer{bdef}
Note that the explicit dependence on $S$ in $\nabla^2
H_{\alpha\beta}$ is canceled by the S-terms in ${\cal
G}_{\alpha\beta\gamma}$ and $\eta_\alpha$. $H_{\alpha\beta}$ is
thus independent of $S$, in agreement with the fact that it
does not transform under the Super-Weyl transformations (\ref{swt}).

The component local supersymmetry transformations that leave 
the component action invariant are
\begin{eqnarray}
\delta A^\Lambda&=&-\epsilon^{\alpha}\chi_{\alpha}^\Lambda, 
\nonumber \\
\delta\chi_{\alpha}^\Lambda&=&\epsilon_{\alpha}\kal{T}^\Lambda
-i\epsilon^{\beta}\hat{\nabla}_{\beta\alpha}A^\Lambda,
\nonumber \\
\delta{\cal T}^\Lambda&=&i\epsilon_{\alpha}\hat{\nabla}^{\alpha\beta}\chi_{\beta}^\Lambda
+\frac{1}{2}S\epsilon^{\alpha}\chi_{\alpha}^\Lambda,
\nonumber \\
\delta e_{\alpha\beta}{}^{\mu\nu} &=&
-2i\epsilon^\g\Psi_{\alpha\beta}{}^\delta e_{\delta\g}{}^{\mu\nu},
\nonumber \\
\delta \Psi_{\alpha\beta}{}^\rho &=& 
{\cal D}_{\alpha\beta}\epsilon^\rho 
-2i\epsilon^\mu\Psi_{\alpha\beta}^{\delta}\Psi_{\mu\delta}^{\rho}
+\frac{i}{2}S\epsilon_{(\alpha}\delta_{\beta)}^\rho,
\nonumber \\
\delta S&=&-\epsilon^{\alpha}\eta_{\alpha},
\nonumber \\
\delta h_{\alpha}&=&-\epsilon_{\alpha}b-\epsilon^{\beta}h_{\beta\alpha},
\nonumber \\
\delta h_{\alpha\beta}&=&-\epsilon^{\gamma}\phi_{\gamma\alpha\beta},
\nonumber \\
\delta b&=&-\epsilon^{\alpha}b_{\alpha}, 
\nonumber \\
\delta\lambda_{\alpha}&=&-\epsilon^{\beta}f_{\beta\alpha},
\nonumber \\
\delta\mu_\a&=&\e^\beta\left[i\kal{D}_{\beta\g}h^\g_{\ \a}+{3\over 2}S
h_{\a \beta}-\kal{D}_{\a \beta}b +i\Psi_{\beta \g}^{\ \ \s}\phi_{\s\
\a}^{\ \g}-i\Psi_{\a\beta}^{\ \ \g}b_\g-G_{\a\beta\g}h^\g\right]\cr
&& +\frac{1}{4}\e_\a(\eta^\g h_\g-2Sb),
\nonumber \\
\delta f_{\alpha\beta}&=&
-\epsilon^{\gamma}\phi_{\gamma\alpha\beta}
-S\epsilon_{(\alpha}\lambda_{\beta)}.
\end{eqnarray}

\section{Component actions}

Having defined the components and derived the relations in \enr{xdef}
-\enr{bdef}, we are in a position to find the component actions from
\enr{sbi1} and
\enr{sbi2} . We use the $3D$ density formula derived in \pek{ulmr}:
\ber
\int d^3\xi d^2\th E^{-1}\kal{L}=\int d^3\xi
e^{-1}\left[\left(\N^2+i\F^{\a\beta}_{\ \ \beta}\N_\a+2S+\F_{\a(\beta}
^{\ \ \ \ \beta}\F_{\g)}^{\ \ \a\g}\right)\kal{L}\right]|.
\eer{defo}

We first want to establish the equivalence of the bosonic parts of
\enr{sbi1} and
\enr{sbi2} to \enr{eha}. The bosonic contents of the actions simply follow from
taking the bosonic part of the first term on the right hand side of
\enr{defo}\footnote{Since the pure bose part of ${\cal L}$ is zero,
the $S$-term will not contribute.}.
We thus find for the action in \enr{sbi1}
\beq
I_1\to \int d^3\xi {e^{-1}\over\sqrt{1-h^2}}\left[
\Omega-2\kal{T}^2\right]\Omega^{\frac{1}{2}},
\eeq{i1b}
where $\Omega$ is the bosonic part of $\kal{S}$, defined in \enr{fdef}, i.e., 
it is the
factor which is raised to $3/2$ in the action \enr{eha}.
Obviously, integrating out the auxiliary field $\kal{T}$, we recover the action
\enr{eha}.

The action in \enr{sbi2} is a little more challenging. It gives, (keeping the
spinor notation),
\ber
&&I_2\to \cr
&&-\int d^3\xi {e^{-1}\over\sqrt{1-h^2}}\left[
\frak{1}{2}\Omega^2+\left(-i\kal{T}\dott{\cal D}_{\a\beta}A+\half 
h_{\ (\a}^\s f_{\beta)\s}\right)\left(-i\kal{T}\dott{\cal D}^{\a\beta}A+\half 
h^{\g(\a} f^{\beta)}_{\ \ \g}\right)\right]\Omega^{-1/2},\cr
&&\ \
\eer{i2b}
where dot denotes contraction over the ambient spacetime indices.
It is a remarkable
fact, manifest in \enr{i2b}, that integrating out the auxiliary field
$\kal{T}$ we again recover the action \enr{eha}.

From the expressions \enr{i1b} and \enr{i2b} we may also read off the
$\kal{T}$ equations that result from the super-Weyl invariant action $\t
I_1-\frak{2}{3}I_2$, they are
\beq
\Omega\kal{T}^\Lambda+\frak{i}{3}{\cal D}^{\a\beta}A^\Lambda h_\a^{\ \g} f_{\beta\g}
+\frak{1}{3}{\cal D}^{\a\beta}A^\Lambda(\kal{T}\dott{\cal D}_{\a\beta}A)=0,
\eeq{swg}
which we only need to solve for $\kal{T}\dott{\cal D}_{\a\beta}A$, where
\beq
(\kal{T}\dott{\cal D}_{\a\beta}A)\kal{M}_{\g\de}^{\a\beta}={-i\over {3\Omega}}
({\cal D}_{\g\de}A\dott{\cal D}^{\a\beta}A)h_\a^{\ \g} f_{\beta\g},
\eeq{swg2}
and
\beq
\kal{M}_{\g\de}^{\a\beta}\is \frak{1}{2}\de_{(\g}^\a\de_{\de )}^\beta
+{1\over {3\Omega}}({\cal D}_{\g\de}A\dott{\cal D}^{\a\beta}A).
\eeq{me}

We note that for $W_\a=0$ the fact that $\kal{M}$ is non-degenerate ensures
that $\kal{T}^\Lambda=0$ and thus that the correct bosonic action results in that
case.
For $W_\a\ne 0$ the equation \enr{swg2} is still solvable, the solution being
expressed in the inverse $\kal{M}^{-1}$ of the three by three matrix. It is
clear, though, that the resulting action will be rather complicated and
contain unwanted bosonic terms. For this reason we  have not 
calculated it explicitly. 

The component actions are
\ber
\label{I1comp}
I_{1}&=&\int d^3\xi {e^{-1}\over\sqrt{1-h^2}}\left[
\Omega^{\frac{3}{2}} \right.\cr
&&\left. \ \ \ \ +\Omega^{\frac{1}{2}}
\{2i\chi^{\alpha} \dott \hat{\nabla}_{\beta\alpha}\chi^{\beta}
+\frac{1}{2}\sigma^{\gamma}_{\ \gamma\alpha}h^{\alpha}
-2{\cal T}^{2}
-\mu^\alpha\lambda_\alpha 
+P\Psi_{\alpha(\beta}^{\ \ \ \ \beta}\Psi_{\gamma)}^{\ \ \alpha\gamma}\right.\cr
&&\left. \ \ \ \ \ \ \ \ \ \ \ \ \ +i\Psi^{\alpha\beta}_{\ \ \ \beta}
(-2i\zeta_{\alpha}-2{\cal T} \dott \chi_{\alpha}
-h_{\alpha\gamma}\lambda^{\gamma}-b\lambda_{\alpha}
+h^{\gamma}f_{\alpha\gamma})
+SP \}
\right.\cr
&&\left. \ \ \ \ +\Omega^{-\frac{1}{2}}
\{\frac{1}{2}
(2i\zeta_{\gamma}+2{\cal T} \dott \chi_{\gamma}+h_{\gamma\alpha}\lambda^{\alpha}
+b\lambda_{\gamma}-h^{\alpha}f_{\gamma\alpha})\xi^{\gamma}+
\right.\cr
&&\left. \ \ \ \ \ \ \ \ \ \ \ \ \ \
\frac{1}{4}P[4(
\hat{\nabla}_{\delta\epsilon}{\cal T} \dott \hat{\nabla}^{\delta\epsilon}A
-\frac{1}{2}\hat{\nabla}_{\delta\epsilon}\chi_{\gamma} \dott 
  \hat{\nabla}^{\delta\epsilon}\chi^{\gamma}
-i\chi^{\sigma} \dott \hat{\nabla}^{\delta\epsilon}A
{\cal G}_{\delta\epsilon\sigma})
\right.\cr
&&\left. \ \ \ \ \ \ \ \ \ \ \ \ \ \ 
+\frac{8i}{3}\hat{\nabla}^{\delta\epsilon}A \dott \chi_{(\delta}\eta_{\epsilon )}
+2\phi_{\delta\epsilon}f^{\delta\epsilon}
-2\phi_{\gamma\delta\epsilon}\sigma^{\gamma\delta\epsilon}
+2h_{\delta\epsilon}\sigma^{\delta\epsilon}
-2S\Psi_{\delta\epsilon}^{\ \ \rho}\chi_{\rho} \dott \hat{\nabla}^{\delta\epsilon}A
\right.\cr
&&\left. \ \ \ \ \ \ \ \ \ \ \ \ \ \ 
+4iS\chi^{\beta} \dott \hat{\nabla}_{\gamma\beta}\chi^{\gamma}
+2iSh_{\delta\epsilon}( 
\Psi_\rho{}^{\delta\epsilon} \lambda^\rho - \Psi^{\rho\delta}{}_\rho \lambda^\epsilon)
+4S\phi^{\gamma}_{\ \gamma\beta}\lambda^{\beta}
\right.\cr
&&\left. \ \ \ \ \ \ \ \ \ \ \ \ \ \
-3S^{2}\chi^2
+2i\Psi^{\gamma\beta}{}_\beta\xi_{\gamma}]\}
\right.\cr
&&\left. \ \ \ \ -\frac{1}{8}\Omega^{-\frac{3}{2}}
P\xi^2 \right.\cr
&&\left. \ \ \ 
-\frac{1}{(1-h^2)}\{\Omega^{\frac{1}{2}}
\{\frac{1}{2}P
[\phi_{\gamma\delta\epsilon}\phi^{\gamma\delta\epsilon}
-2h_{\delta\epsilon}\phi^{\delta\epsilon}
-2i\Psi^{\delta\epsilon}_{\ \ \epsilon}
\phi_{\delta\rho\sigma}h^{\rho\sigma}]
\right. \cr
&&\left. \ \ \ \ \ \ \ \ \ \ \ \ \ \ \ \ \ \ \ \ \ \ \ \ \ 
+\phi^\gamma{}_{\alpha\beta}h^{\alpha\beta}
(2i\zeta_{\gamma}+2{\cal T} \dott \chi_{\gamma}
+h_{\gamma\delta}\lambda^{\delta}+b\lambda_{\gamma}
-h^{\delta}f_{\gamma\delta}\}
\right.\cr 
&&\left. 
\ \ \ \ \ \ \ \ \ \ \ \ \ \ \ \ 
+\frac{1}{2}\Omega^{-\frac{1}{2}}
P\phi_{\gamma\alpha\beta}h^{\alpha\beta}\xi^{\gamma}\}
\right.\cr 
&&\left. \ \ \ +\frac{3}{2(1-h^2)^{2}}
\Omega^{\frac{1}{2}}
P
\phi^{\gamma}_{\ \delta\epsilon}h^{\delta\epsilon}
\phi_{\gamma\sigma\rho}h^{\sigma\rho}
\right]  
\eer{35}
and
\ber
\label{I2comp}
I_{2}&=&\int d^3\xi {e^{-1}\over\sqrt{1-h^2}}\left[
-\half \Omega^\frac{3}{2} \right. \cr
&&\left. \ \ \ \ \ \ \ \ \ \ \ \
+\Omega^\half \{
[i{\cal T}_\gamma{}^\gamma + \frac{3}{2}S\chi^2
-\phi_\gamma{}^{\gamma\alpha}\lambda_\alpha ] 
+\Psi_\gamma{}^\beta{}_\beta\zeta^\gamma\} \right. \cr
&&\left. \ \ \ \ \ \ \ \ \ \ \ \
+\Omega^{-\frac{1}{2}}
\{-[2\hat{\nabla}^{\gamma\alpha}\chi_{\gamma} \dott \hat{\nabla}_{\alpha\beta}A
+2\hat{\nabla}^{\gamma\alpha}A \dott \hat{\nabla}_{\alpha\beta}\chi_{\gamma}
+3S{\cal T} \dott \chi_{\beta}+2i\chi^{\alpha} \dott \hat{\nabla}_{\alpha\beta}{\cal T}
\right.\cr
&&\left. \ \ \ \ \ \ \ \ \ \ \ \ \ \ \ \ \ \ \ \ \ \ \ \ \ 
-2i{\cal T} \dott \hat{\nabla}_{\alpha\beta}\chi^{\alpha}
-2\chi^2\eta_\beta
-\half iS\chi^2\Psi^\alpha{}_{\beta\alpha}
\right. \cr
&&\left. \ \ \ \ \ \ \ \ \ \ \ \ \ \ \ \ \ \ \ \ \ \ \ \ \
+2\phi_{\beta}^{\ \alpha}\lambda_{\alpha}
-2\phi_{\gamma\beta}{}^{\alpha}f^{\gamma}_{\ \alpha}
+h_{\beta}^{\ \alpha}\sigma^{\gamma}_{\ \gamma\alpha}
-3Sh_{\beta}^{\ \alpha}\lambda_{\alpha}]i\zeta^{\beta}
\right.\cr
&&\left. \ \ \ \ \ \ \ \ \ \ \ \ \ \
-(i{\cal T}^{\gamma\beta}
+\frac{3}{4}SC^{\gamma\beta}\chi^2
-\phi^{\gamma\beta\delta}\lambda_{\delta})
(i{\cal T}_{\gamma\beta}
+\frac{3}{4}SC_{\gamma\beta}\chi^2
-\phi_{\gamma\beta}^{\ \ \alpha}\lambda_{\alpha})\right.\cr
&&\left. \ \ \ \ \ \ \ \ \ \ \ \ \ \
+2\Psi^{\alpha\beta}_{\ \ \beta}
(-i{\cal T}_{\alpha\gamma}-\frac{3}{4}SC_{\alpha\gamma}\chi^2
+\phi_{\alpha\gamma}^{\ \ \sigma}\lambda_{\sigma})
\zeta^{\gamma}
\right.\cr
&&\left. \ \ \ \ \ \ \ \ \ \ \ \ \ \
-\zeta^2
(2S+\Psi_{\alpha(\beta}{}^{\beta}\Psi_{\gamma)}^{\ \alpha\gamma})+
i\half\zeta_\gamma\xi^\gamma\} 
\right.\cr
&&\left. \ \ \ \ \ \ \ \ +\Omega^{-\frac{3}{2}}
\{(i{\cal T}_{\gamma\alpha}
+\frac{3}{4}SC_{\gamma\alpha}\chi^2
-\phi_{\gamma\alpha}^{\delta}\lambda_{\delta})
i\zeta^{\alpha}\xi^{\gamma}
\right.\cr
&&\left. \ \ \ \ \ \ \ \ \ \ \ \ \ \
+\frac{1}{4}\zeta^2
\left\{ 4 \left[\hat{\nabla}_{\delta\epsilon}{\cal T} \dott
\hat{\nabla}^{\delta\epsilon}A
-\frac{1}{2}\hat{\nabla}_{\delta\epsilon}\chi_{\gamma} \dott
\hat{\nabla}^{\delta\epsilon}\chi^{\gamma}
-i\chi^{\sigma} \dott \hat{\nabla}^{\delta\epsilon}A
{\cal G}_{\delta\epsilon\sigma}\right]
\right. \right. \cr
&&\left. \ \ \ \ \ \ \ \ \ \ \ \ \ \ \ \ \ \ 
+\frac{8i}{3}\hat{\nabla}^{\delta\epsilon}A \dott \chi_{(\delta}\eta_{\epsilon )}
+2\phi_{\delta\epsilon}f^{\delta\epsilon}
-2\phi_{\gamma\delta\epsilon}\sigma^{\gamma\delta\epsilon}
+2h_{\delta\epsilon}\sigma^{\delta\epsilon}
-2S\Psi_{\delta\epsilon}^{\ \ \rho}\chi_{\rho} \dott \hat{\nabla}^{\delta\epsilon}A
\right.\cr
&&\left. \ \ \ \ \ \ \ \ \ \ \ \ \ \ \ \ \ \ 
+4iS\chi^{\beta} \dott \hat{\nabla}_{\gamma\beta}\chi^{\gamma} 
+2iSh_{\delta\epsilon}( 
\Psi_\rho{}^{\delta\epsilon} \lambda^\rho - \Psi^{\rho\delta}{}_\rho \lambda^\epsilon)
+4S\phi^\gamma{}_{\gamma\beta}\lambda^\beta
\right.\cr
&&\left. \left. \ \ \ \ \ \ \ \ \ \ \ \ \ \ \ \ \ \
-3S^{2}\chi^2 
+2i\Psi^{\alpha\beta}_{\ \ \beta}
\xi_{\alpha}\right\}\} 
-\frac{3}{8}\Omega^{-\frac{5}{2}} \zeta^2 \xi^2
\right.\cr
&&\left. \ \ \ -\frac{1}{(1-h^2)}\{i\Omega^\half
\phi_{\alpha\lambda\sigma}h^{\lambda\sigma}\zeta^\alpha
\right. \cr
&&\left. \ \ \ \ \ \ \ \ \
+\Omega^{-\frac{1}{2}}
\{-\frac{1}{2}\zeta^2
(\phi_{\gamma\delta\epsilon}\phi^{\gamma\delta\epsilon}
-2h_{\delta\epsilon}\phi^{\delta\epsilon})
\right.\cr
&&\left.  \ \ \ \ \ \ \ \ \ \ \ \ \ \ \ \ \ \ \ 
-2\phi_{\gamma\lambda\sigma}h^{\lambda\sigma}
(-i{\cal T}^{\gamma}_{\ \alpha}
-\frac{3}{4}SC^{\gamma}_{\ \alpha}\chi^2
+\phi^{\gamma\ \ \delta}_{\ \alpha}\lambda_{\delta})
i\zeta^{\alpha}
\right.\cr
&&\left.  \ \ \ \ \ \ \ \ \ \ \ \ \ \ \ \ \ \ \
+i\Psi^{\alpha\beta}_{\ \beta}
\zeta^2
\phi_{\alpha\sigma\rho}h^{\sigma\rho}\} \right.\cr
&&\left. \ \ \ \ \ \ \ \ 
+\Omega^{-\frac{3}{2}}\frac{1}{2}
\zeta^2
\phi_{\gamma\sigma\rho}h^{\sigma\rho}
\xi^{\gamma}\} \right.\cr
&&\left. \ \ \ -\frac{3}{2(1-h^2)^{2}}\Omega^{-\frac{1}{2}}
\zeta^2
\phi^{\gamma}_{\ \delta\epsilon}h^{\delta\epsilon}
\phi_{\gamma\sigma\rho}h^{\sigma\rho} \right]  
\eer{36}
where we have introduced the definitions
\begin{eqnarray}
  \Omega&\equiv & \hat{\nabla}_{\alpha\beta}A \dott \hat{\nabla}^{\alpha\beta}A
        +h_{\alpha\beta}f^{\alpha\beta} \nonumber \\
  P &\equiv& \chi^2+h_\alpha\lambda^\alpha \nonumber \\
  \zeta_{\alpha}&\equiv& \chi^{\beta} \dott \hat{\nabla}_{\alpha\beta}A+
        ih_{\alpha\beta}\lambda^\beta \nonumber \\
  \xi_{\gamma}&\equiv &2\hat{\nabla}^{\alpha\beta}A \dott
        \hat{\nabla}_{\alpha\beta}\chi_{\gamma} +
        \phi_{\gamma\alpha\beta}f^{\alpha\beta}+
        \sigma_{\gamma\alpha\beta}h^{\alpha\beta}-2iS\zeta_\gamma \nonumber \\
  {\cal T}_{\alpha\beta}&\equiv&{\cal T} \dott \hat{\nabla}_{\alpha\beta}A
       +\chi^{\gamma} \dott \hat{\nabla}_{\beta\gamma}\chi_{\alpha}+
       \frac{i}{2}h_{(\alpha}{}^\gamma f_{\beta)\gamma} \nonumber \\
  \chi^2 &\equiv& \chi^\alpha \dott \chi_\alpha \ \ \ \
  \zeta^2 \equiv \zeta^\alpha \zeta_\alpha \ \ \ \
  \xi^2 \equiv \xi^\alpha \xi_\alpha \ \ \ \
  {\cal T}^2 \equiv {\cal T} \dott {\cal T}.
\end{eqnarray}

The component version of the super-Weyl invariant combination \enr{swa} can
be reassembled from
\enr{35} and \enr{36} (omitting the terms from $H^\a W_\a$). The
super-Weyl invariance should manifest itself in an independence of
$S$. We have only checked the quadratic $S$-terms which indeed cancel.

\section{Discussions and conclusions}

We have studied a first order ``Weyl invariant'' bosonic action for
$Dp$-branes for the special case of $p=2$ and shown how it can be coupled
to $3D$ supergravity in (at least) two different ways. We have further
derived the component content of the model, both for the actions and the
local supersymmetry transformations. Our results generalize those for the
spinning membrane, except when it comes to super-Weyl invariance. The
super-Weyl invariant action we find does indeed reduce to that of the
spinning membrane when we turn off the world-volume Maxwell fields, but it
does {\it not}  have the correct bosonic limit when they are non-zero.
Although we have not been able to find one, we see no reason why
another super-Weyl invariant action with the correct limit should not
exist. To construct it one would have to have some additional guideline,
however.

An interesting question is of course whether the present construction
generalizes to any other $p$. Many of the features we encountered in the
above constructions are three dimensional. E.g., the fact that we were
able to reduce $s_{ab}\to e_a,h_a$ is based on the $3D$ equivalence of a
vector and an antisymmetric tensor. For higher $p$ we typically have
$h_{ab}$ instead of $h_a$ and we thus expect the potential to have spin
$3/2$ instead of spin $1/2$. The form of the superspace action will have
to change because of the growth of the superspace measure. There is also
the question of which multiplets to use. All in all, it looks
as if each case has to be considered separately.

One of the perhaps unwanted features of the bosonic starting point is the
non-linearity in \enr{sa2}. Usually, going from a Nambu-Goto type action
to an action with an auxiliary metric leads to a quadratic behaviour in the
path integral (after gauge fixing). This is not the case here, although we
replace a square root of a determinant by powers of $\d X$.
Further linearization may formally be achieved by introducing additional auxiliary
fields and lagrange multipliers, but does not really seem very useful.

As far as we know, locally supersymmetric extensions of the Born-Infeld
action have not been discussed previously. However, several globally
(world volume) supersymmetric models in various dimensional superspaces
exist. E.g.,
\pek{depu, cefe, baga, rotse} ($p=3, N=1$) and \pek{rotse, ket1, ket2}
($p=5, N=1$ and $p=3, N=2$). It would be interesting to compare the 
present models with the globally supersymmetric D2-brane, which is 
discussed in \pek{Ivanov:1999fw}. 

As mentioned in the introduction, the
motivation for studying the globally supersymmetric extensions of the
$BI$-action is usually either taken to be its appearence in the effective
action of the
$10D$ open superstring or its role in the $D$-brane effective action. In
both these cases it is also interesting to investigate the possible
existence of locally supersymmetric extensions, and the present results is
a first contribution to this.

\begin{flushleft} {\bf Acknowledgments}
\end{flushleft}

\noindent We thank Martin Ro\v cek for inspiring discussions that
initiated this work. We are also grateful to Bo Sundborg for comments.
The work of UL was supported by the Swedish Research Council under 
contract F-AA/FUO4038-312 and by NorFA under contract 9660003-0.


\newpage

\end{document}